\documentclass[11pt,preprint,floatfix]{revtex4}
\usepackage[dvips]{graphicx}
\usepackage{subfigure}
\usepackage{morefloats}
\input{psfig.sty}

\newif\ifpdf
\ifx\pdfoutput\undefined
\pdffalse 
\else
\pdfoutput=1 
\pdftrue
\fi
\ifpdf
\else
\fi
\begin{document}

\title{Scaling in Non-stationary Time Series II: Teen Birth Phenomenon}
\author{M. Ignaccolo$^{1}$\footnote{Corresponding Author.\newline \textit{Mailing Address}: Center for Nonlinear Science, University of North Texas,
P.O. Box 311427, Denton, Texas, 76203-1427 . \newline \textit{Phone}: +1 940 565 3280 . \newline \textit{E-mail Address}: stellina99@earthlink.net . }, P. Allegrini$^{2}$, P. Grigolini$^{1,3,4}$, P. Hamilton$^{5}$, B. J. West$^{6}$}
\address{$^{1}$Center for Nonlinear Science, University of North Texas,
P.O. Box 311427, Denton, Texas, 76203-1427} 
\address{$^{2}$ Istituto di Linguistica Computazionale del Consiglio Nazionale delle
Ricerche,\\
 Area della Ricerca di Pisa-S. Cataldo, Via Moruzzi 1, 
56124, Ghezzano-Pisa, Italy } 
\address{$^{3}$Dipartimento di Fisica dell'Universit\`{a} di Pisa and INFM 
Via Buonarroti 2, 56127 Pisa, Italy } 
\address{$^{4}$Istituto di Biofisica del Consiglio Nazionale delle
Ricerche,\\ Area della Ricerca di Pisa-S. Cataldo, Via Moruzzi 1,
56124, Ghezzano-Pisa, Italy } 
\address{$^{5}$ Center for Nonlinear Science, Texas Woman's University, 
P.O. Box 425498, Denton, Texas 76204} 
\address{$^{6}$ Physics Department,
Duke University, P.O. Box 90291, Durham, North Carolina 27708
and US Army Research Office, 
Mathematics Division, 
Research Triangle Park, NC 27709}
\date{\today}

\begin{abstract}

This paper is devoted to the problem of statistical mechanics raised by the 
analysis of an issue of sociological interest: the teen birth
phenomenon. It is expected that these data are characterized by correlated
fluctuations, reflecting the cooperative properties of the process. However,
the assessment of the anomalous scaling generated by these correlations is
made difficult, and ambiguous as well, by the non-stationary nature of the
data that show a clear dependence on seasonal periodicity (periodic component) and an average changing 
slowly in time (slow component), as well. We use the detrending techniques described in the companion paper \cite{paper1}, 
to safely remove all the biases and to derive the genuine scaling of the teen birth phenomenon.
\textit{keywords}: sociological time series, complexity, scaling, detrending methods 

\end{abstract}

\maketitle

\ifpdf
\DeclareGraphicsExtensions{.pdf, .jpg, .tif} \else
\DeclareGraphicsExtensions{.eps, .jpg} \fi

\section{Introduction}\label{intro} 

The difficulty in obtaining a stable and reliable measure of
complexity is one of the main problems of time series analysis. Since the
pioneer work of Ref. \cite{stanley} the measure of complexity has been
determined having in mind Brownian motion as a condition of uncorrelated
randomness. Consequently we expect a significant deviation from Brownian
motion when we analyze time series characterizing complex systems. This deviation reflects the cooperative
character of interactions between the different constituents of a complex system. The analysis of this deviation is carried out by using
 the time series fluctuations to generate a diffusion process and the diffusion
variable $x$ is expected to deviate from the condition of ordinary Brownian
motion. If the time series is sufficiently long, the probability
distribution function (pdf) of the resulting diffusion process, $p(x,t)$, is
expected to satisfy the scaling condition

\begin{equation}
p(x,t)=\frac{1}{t^{\delta }}F(\frac{x}{t^{\delta }}).
\label{scalingdefinition}
\end{equation}

The sociological phenomenon of teen births studied in this paper yields a
time series of daily counts of births to women less than $20$ years old in a span of $36$ years (from $1964$ to $1999$), 
to which we shall refer, from now on, as the ``teen birth data". The 
influence of an external bias is likely to be significant in such data.
The analysis of such non-stationary time series encounters the risk of
producing the misleading impression that the underlying process is complex,
with no real connection between the complexity measure and the cooperation
among society's components. One possible source of ambiguity in the
complexity measures is the \emph{annual periodicity} in the number of births each day. 
We do not rule out the possibility that such a periodicity might have a social origin, rather than
being tied to seasonal changes. However, such periodicity might also depend
on biological effects. Yet, if not properly taken into account, the annual
periodicity might give the illusion of strong cooperation that does not
exist in the dynamics of the process. Another possible misleading source of
complexity measure is the increase in the daily birth rate over time. We do
not rule out the possibility that this increase might have a complex social
origin, either. However, this increase might also originate from an overall
increase in the underlying population and not be a property of the cohort
group at all. We refer to such an increase as \emph{demographic pressure}. 

In other words, the perspective that we adopt in this paper is the
following. We imagine that the time series consists of erratic changes in
the number of births embedded in an increasing bias and an annual
periodicity. We imagine that both the bias and perodicity are external to
the process under study. This separation is assumed for clarity and
convenience, since we do not have any evidence that these properties are
dependent on the internal dynamics of the process under study. We want to
establish that annual cycles and demographic pressure need not have anything
to do with cooperation, and system complexity, even if we do not rule out
the possibility that both of them might reflect important sociological
influences on the system under study. We intend to show that the only
unquestionable measure of complexity that remains after detrending the
annual cycles and demographic pressure is the correlation.

In Section \ref{tb} we introduce the teen birth data from the State of Texas and discuss some of the time series most obvious
properties. In Sections \ref{slcomp} and \ref{percomp}, using the results of \cite{paper1}, 
we eliminate from the teen birth data, the influences of the bias due to the demographic pressure and of the yearly and weekly periodicities, respectively. In Section \ref{ultimategoal} we apply the diffusion entropy analysis to the remaining signal, that is, to the fluctuations with the bias due to the demographic pressure, and
periodicities removed. The residual scaling in these fluctuations is
interpreted in terms of the intrinsic dynamics of the system. In Section \ref{conclusion}
we draw some conclusions.

\section{Texas Teen Birth Data}\label{tb} 

\begin{figure}[tbp]
\includegraphics[angle=270,width=5.5in] {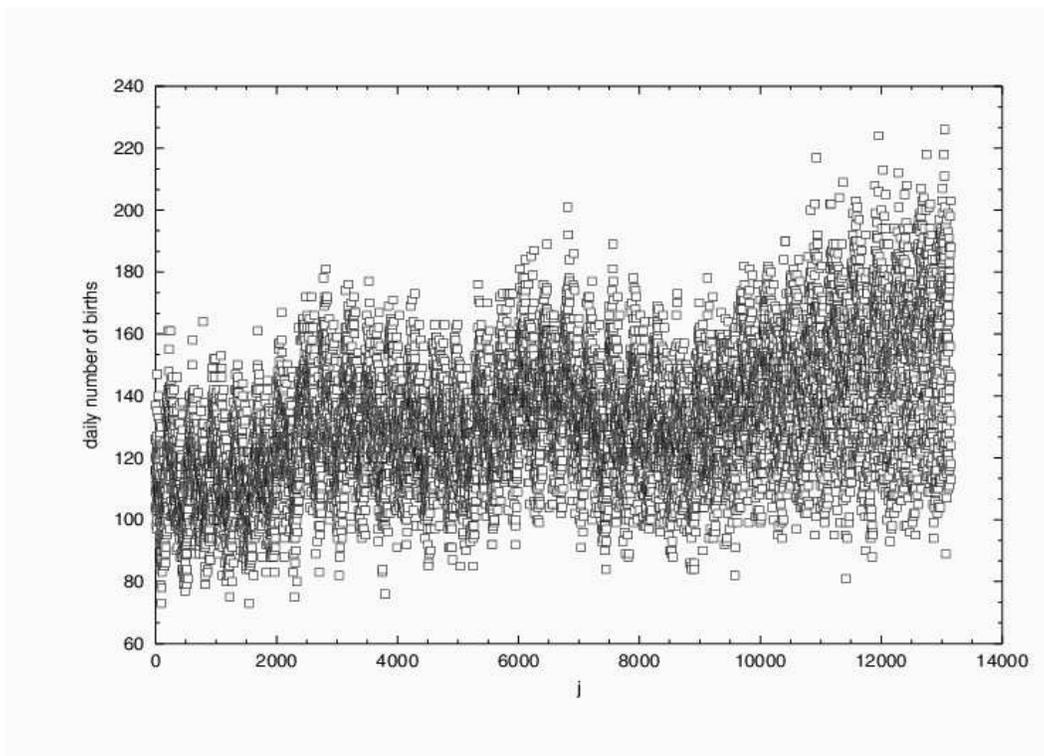}
\caption{The daily number of births from a teenager mother in the whole state of Teaxs, from the $1$st January $1964$ ($j=1$) to the $31$ st  December $1999$ ($j=13149$).}
\label{teenbirthdata}
\end{figure}

\begin{figure}[tbp]
\includegraphics[angle=-90,width=5.5in] {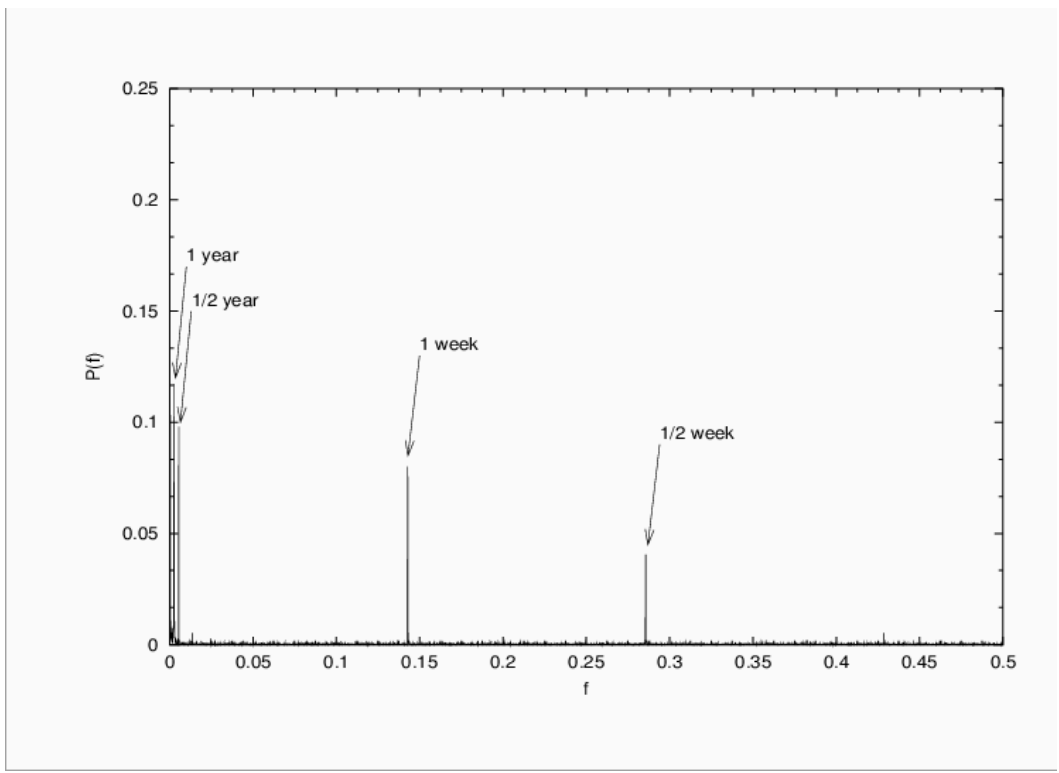}
\caption{The power $P(f)$ as a function of the frequency $f$ for the teen
birth data of Fig. 1.}
\label{powerspectrum}
\end{figure}

The teen birth data, the daily number of births to teenager mothers in the whole state
of Texas from 1964 to 1999, are illustrated  in Fig. \ref{teenbirthdata}.
A closer look to the data reveals immediately an annual
periodicity. It has been pointed out \cite{patti} that this kind of
periodicity is related to the natural annual cycle of a woman's fertility, here we limit ourselves to examining this periodic process with
the methods of statistical mechanics. First of all, we apply to the signal
a power spectrum analysis. The results of this analysis are illustrated in
Fig. \ref{powerspectrum}. We see clearly that in addition to the expected
annual periodicity, there are also exist a six month, a one week and a
half-week component. The weekly periodicity can be related to the practice
of scheduling inductions and cesarean sections during the week in order to have fewer births on the
weekend. The other important feature of the teen birth data, the demographic pressure, is evident as well. In fact, looking at Fig. \ref{teenbirthdata}, we see how the annual average of births increases, during the $36$
year span considered. For example, starting from the year 1989, $j>9000$ in Fig. \ref{teenbirthdata},  there is a steady rise (on average) in the daily count of births to teens . 

On the basis of these remarks, a reasonable attempt at describing the teen birth data is based on
the assumption that the numbers of births per day $\{\xi _{j}\}$ are given
by the following expression:

\begin{equation}
\xi _{j}= S + \Phi _{j}^{year}+\Phi _{j}^{week}+S_{j}+\zeta _{j},
\label{teenmodel0}
\end{equation}
where $S$ is the mean value of the data, $\Phi _{j}^{year}$ and $\Phi _{j}^{week}$ are respectively an annual and
a weekly periodic component satisfying the condition of zero mean \cite{zeromean}, $S_{j}$ the bias due to the demographic pressure and $\zeta _{j\text{ }}$is an uncorrelated random
variable with zero mean and fixed variance $\sigma _{\zeta }$.  We make the
assumption that all the components of the right-hand side of Eq. (\ref{teenmodel0}) are independent of one another and we shall address, following \cite{paper1}, the function $S_{j}$ as the "slow component".

The main goals of this paper, are to verify the validity of Eq. (\ref{teenmodel0}) as a model of the teen birth phenomenon and to detect the correlation properties of the fluctuation $\zeta $. Both of these goal are achieved through the application of the detrending procedures discussed in \cite{paper1}. 

\section{Processing the Teen Birth Data}\label{dataprocessing}

\subsection{Detrending the Slow Component}\label{slcomp}

\begin{figure}[tbp]
\includegraphics[angle=-90,width=5.5in] {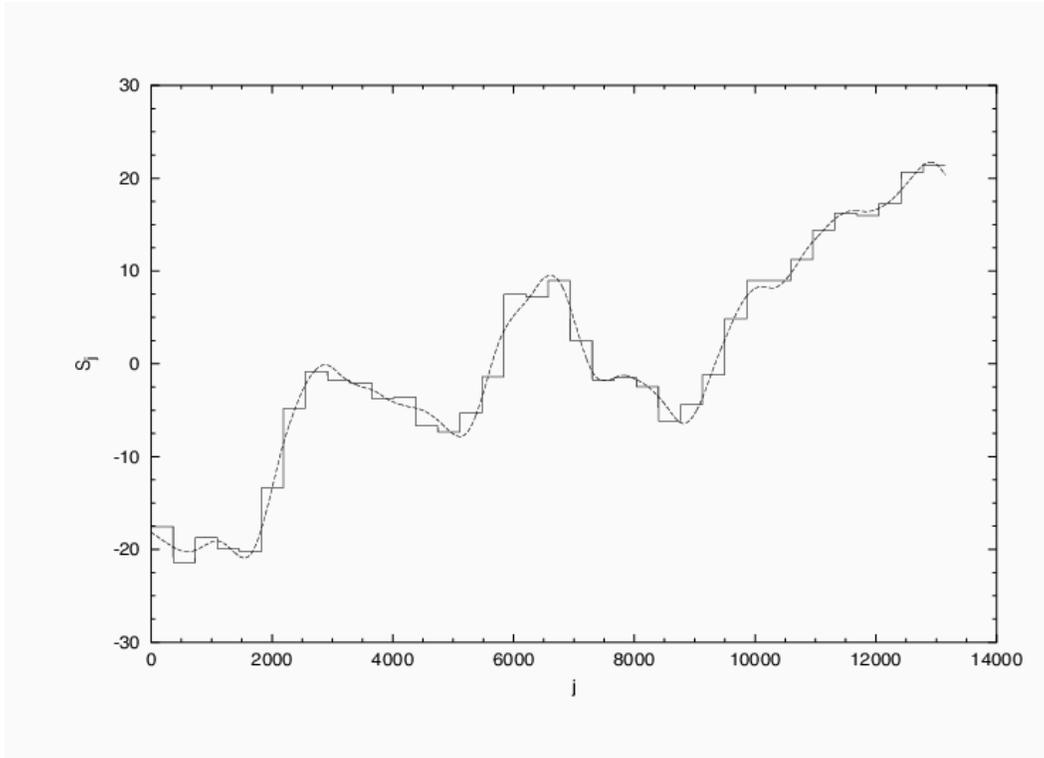}
\caption{The slow component $S_{j}$ as a function of time $j$. The full and
dashed lines denote the result produced by the step smoothing and by the wavelet smoothing, with wavelet
db8, respectively.}
\label{babyslc}
\end{figure}

In \cite{paper1}, we studied the case where we have either a single periodic
component or a single slow component added to noise. In the teen birth data, however,  we have the more complicated situation 
where both biases are superimposed on the random fluctuations in the number of births, and in addition there are two periodic
components rather than one.  Therefore some care must be used in applying the detrending procedures discussed in \cite{paper1}. Let us 
consider, first, the case of a time series where, in addition to a slow component, a periodic one is present. In this case the two 
detrending procedures of \cite{paper1} do not commute: detrending  the periodic component prior to the detrending of
the slow component $S_{j},$ would lead to a distorted periodic bias, while the opposite is possible if the function $S_{j}$ is 
constant for an interval of time equal, at least, to the time period of the
periodic component. With two or more periodic components in the data, the function $S_{j}$ has to be assumed constant for a 
time interval equal to the smallest common multiple of the
time periods of the periodic components. In the case of the teen birth data we have a yearly and a
weekly periodicity added to the slow component $S_{j}$ and considering that a
year is equivalent to ``almost'' $52$ weeks, we can take the length of a year as the smallest common multiple of the two periods 
of these two periodic components. Therefore, following \cite{paper1}, we shall adopt the scale $2^{9}$ when we adopt the wavelet 
detrending procedure (here as in \cite{paper1} we use the Daubechies wavelet number 8 (db8)) and the characteristic time $T=365$ 
for the step detrending procedure. In this last case, using the fact that both $\Phi _{j}^{year}$ and $\Phi _{j}^{week}$ have 
a zero mean, we can write for the sum of the variable $\xi $ inside the j-th patch of length $365$ ($366$ in the case of a leap year)


\begin{equation}
X_{j}=\sum_{k=bj}^{k=bj+365}\xi _{k}\approx \sum_{k=bj}^{k=bj+365}\zeta
_{k}+\sum_{k=bj}^{k=bj+365}S_{k},  \label{sumoveroneyear}
\end{equation}
where $b_{j}$ is the index of the time series relative to the first January
of the j-th (starting from the year 1964) year. Using Eq. (\ref
{sumoveroneyear}) we are able to approximate the slow component $S_{j}$ with
a step function of step length $365$ ($366$ in the case of a  leap year). In Fig. \ref
{babyslc} we plot both the slow components obtained with the wavelet
smoothing and the step smoothing. It is easy see that these are the $%
SS_{j}^{T}$and $SC_{j}^{T}$ that we discussed in \cite{paper1}.

Before proceeding further, an important question to address is whether
we can really assign to the slow component $S_{j\text{ }}$the time scale $%
T=365$, when we adopt the step smoothing, and the time scale $T=2^{9}=512$,
when we adopt the wavelet smoothing. We think that $S_{j}$ has to be
considered fairly smooth because $S_{j}$ represents a sort of daily average
number of the births. This number depends on the daily total teen
population, which, in turns, depends on the number of births occurring approximately $10$ to $19$ years earlier, and so on. 
Moreover we have to take into account
social factors such as immigration or a change of public policy, a Welfare Reform that is aimed at changing reproductive behavior. 
We think, therefore,
that it is plausible to assume the time scale relative to the interplay of
all these factors to be of the order of one year.

\begin{figure}[tbp]
\includegraphics[angle=-90,width=5.5in] {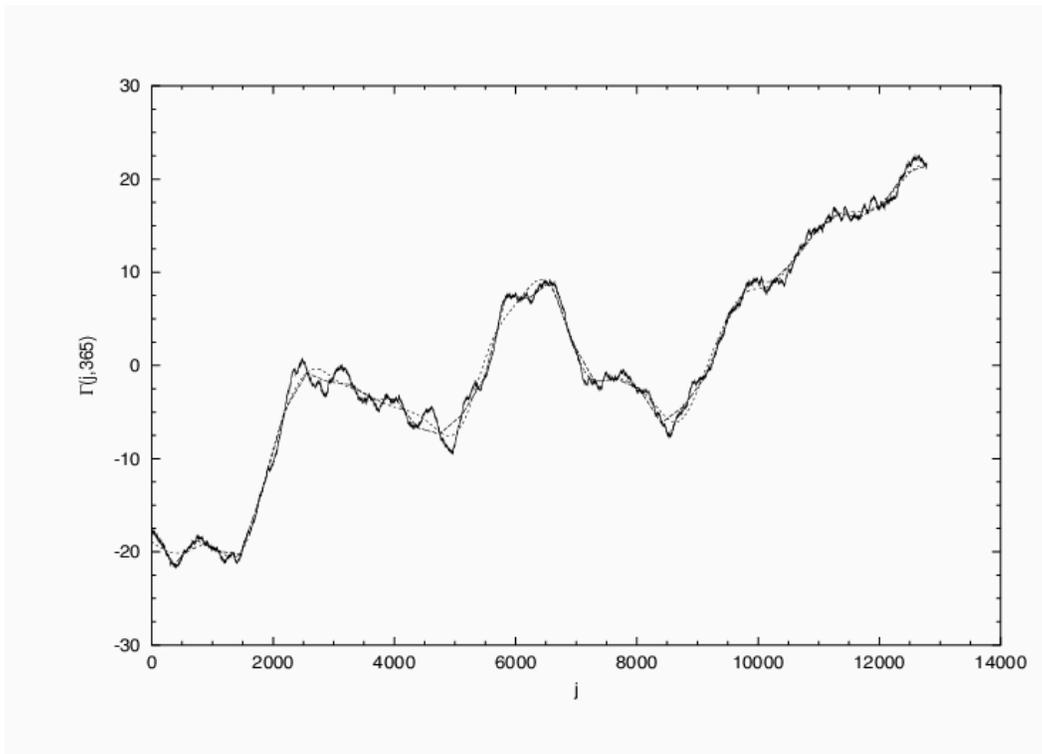}
\caption{The annual moving average of Eq.(\ref{annmovingaverage}). The full
line indicate the result of the numerical analysis applied to the teen birth data, as
they are. The dashed line denotes the the annual moving average applied to
the slow component determined by the step smoothing method. The dotted line
denotes the the annual moving average applied to the slow component
determined by the wavelet smoothing.}
\label{annualmovavg}
\end{figure}

To strengthen these arguments we study the annual moving average. The annual
moving average is defined, \cite{forgetaverage}, by  
\begin{equation}
\Gamma (j,365)\equiv \frac{1}{365}\sum_{k=j}^{k=j+365-1}\xi _{k}\approx 
\frac{1}{365}\sum_{k=j}^{k=j+365-1}S_{k}+\frac{1}{365}\sum_{k=j}^{k=j+365-1}%
\zeta _{k},  \label{annmovingaverage}
\end{equation}
where $j$ goes from $1$ to $(13149-365+1)$. Using the same arguments as in \cite{paper1}, we write
\begin{equation}
\Gamma (j,365)=\frac{1}{365}\sum_{k=j}^{k=j+365-1}S_{k}.
\label{annmovingaverage2}
\end{equation}
where we assume that the intensity of the second contribution to the annual
moving window is given by 
\begin{equation}
\sigma _{j}\times 365^{\delta _{sm}-1}\ll \frac{1}{365}%
\sum_{k=j}^{k=j+365-1}S_{k}.  \label{annmovingaverage3}
\end{equation}
The expression  Eq. (\ref{annmovingaverage2}) for the annual moving average $\Gamma (j,365)$ affords interesting, though indirect, information on the
behavior of function $S_{j}$. In the top frame of Fig. \ref{annualmovavg} we
plot the function $\Gamma (j,365)$. We see that this function is not
dominated by the noise (it is a smooth function). Therefore the
approximation leading to Eq. (\ref{annmovingaverage2}) is a good one. In the
bottom frame of Fig. \ref{annualmovavg} we compare the annual moving average
of the data with the corresponding quantities provided by the slow
components determined by both the step and the wavelet smoothing. We notice
that both methods yield results in good agreement with the the annual moving
average applied to the real data. Therefore, we are encorouged to think that
this slow component is a real property of the data that ought to be
detrended in order to determine the genuine complexity of the time series.

\subsection{Detrending the Periodic Components}\label{percomp}

\begin{figure}[tbp]
\includegraphics[angle=-90,width=5.5in] {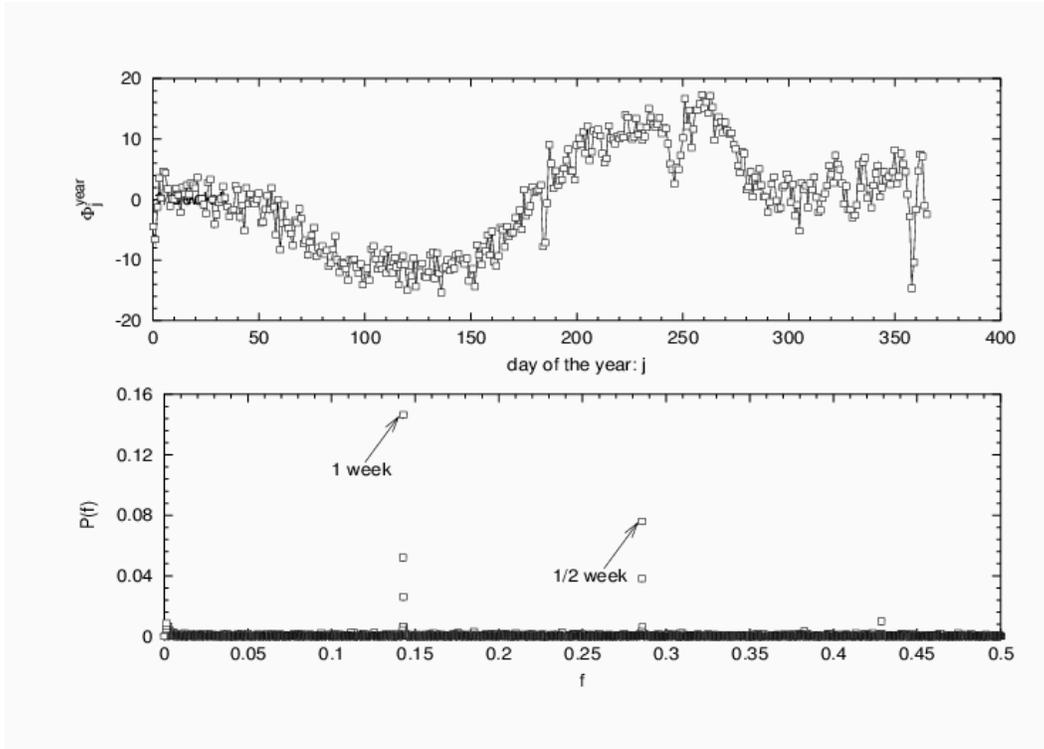}
\caption{Top frame: the annual periodicity $\Phi^{year}_{j}$ of the teen
birth data. The squares indicate the result of the wavelet detrending method, while the
full line denotes the result of the step smothing method. Bottom frame: the
power spectrum of the signal after detrending the yearly periodicity has
been detrended. The squares and the triangles indicate the results of the
wavelet detrending method and of the step smothing method, respectively.}
\label{yearperiodicity}
\end{figure}

After detrending the slow component we can proceed to detrending the two
periodic components. As in the case of the arguments yielding Eqs. (\ref
{sumoveroneyear}) and (\ref{annmovingaverage}), the fact that one year is
almost a multiple of one week simplifies our efforts. In such a case, the periodic component of 
the bigger period can be detrended using the procedure described in \cite{paper1}, leaving 
the periodic component of smaller period unaffected. In the top frame of Fig. \ref{yearperiodicity} we plot the
evaluated yearly periodicity. We notice a sudden drop in the number of
births in correspondence with specific holidays, including the 4th of July ($j=186$), the 1st of
September ($j=245$) and the 24th and 25th of December ($j=359$ and $j=360$).
We also notice no appreciable difference between the results of the two
different recipes, used to detrend the slow component, in both the top and
the bottom frame. The bottom frame shows the power spectrum after detrending
the yearly component, with only the weekly periodicity remaining.

\begin{figure}[tbp]
\includegraphics[angle=-90,width=5.5in] {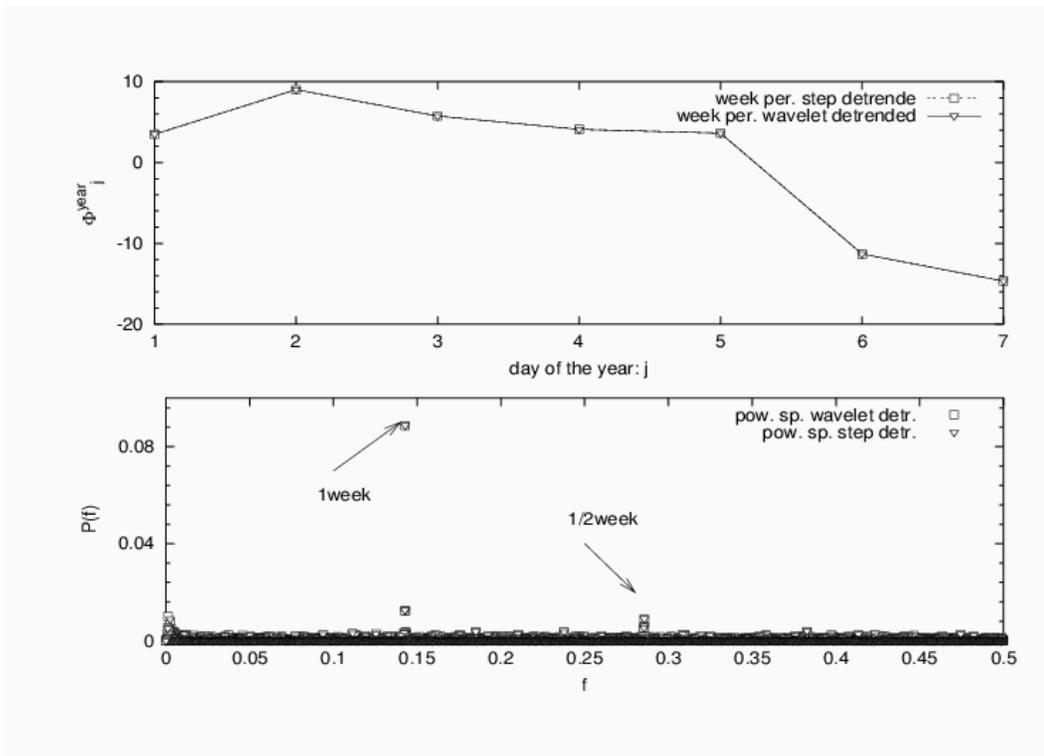}
\caption{Top frame: the week periodicity $\Phi^{week}_{j}$ as a function of the days of
the week, from Monday, $1$ to Sunday $7$. Bottom frame: the power spectrum,
after detrending the week periodicity. In both frames the squares and the
triangle denote the result obtained using the step smoothing procedure and
the wavelet method, respectively.}
\label{weekperiodicity}
\end{figure}

\begin{figure}[tbp]
\includegraphics[angle=-90,width=5.5in] {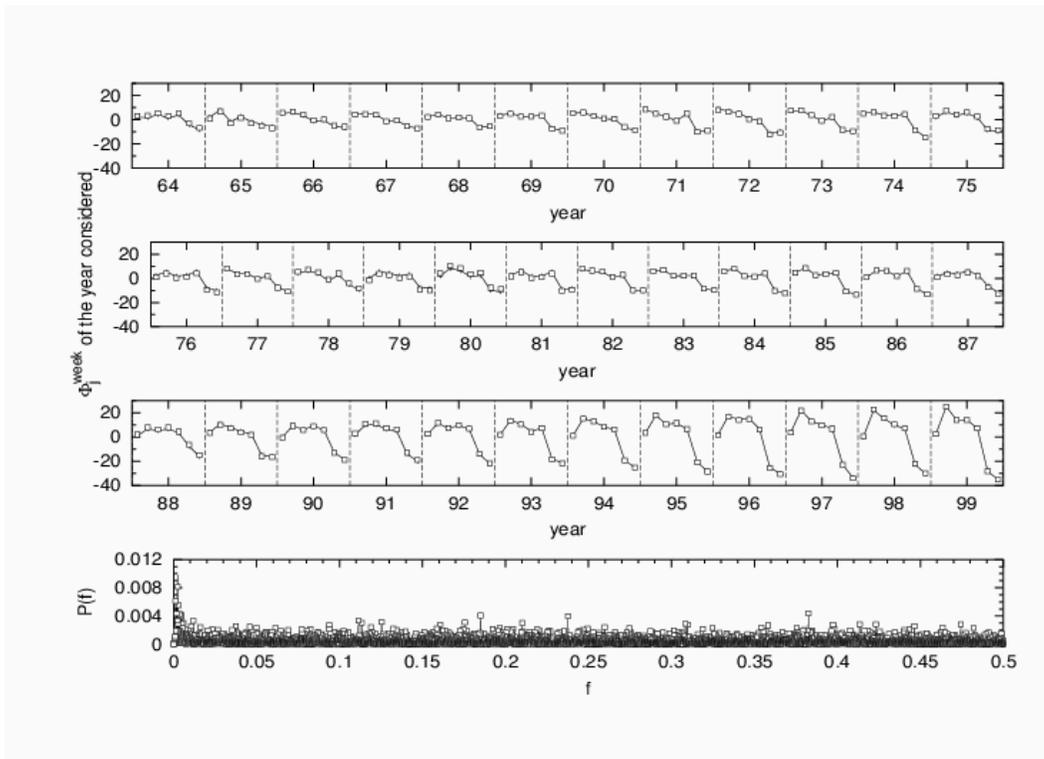}
\caption{Effect of detrending the weekly periodicity year by year. The three top
frames illustrate how the weekly periodicity changes overthe years, from
1964 to 1999. For each year we report seven values corresponding to the
seven days of the week, from Monday to Sunday. The squares and the full line
denote the results stemming from the step detrending method and from the
wavelet decomposition, respectively. The bottom frame is the power spectrum
of the detrended data.}
\label{weekperiodicity2}
\end{figure}

We are now, ready to detrend the weekly periodicity. In Fig. \ref
{weekperiodicity} we ilustrate the corresponding results. The top frame
shows the evaluated weekly periodicity, with the number $1$ representing
Monday, $2$ Tuesday, and so on. As expected, on the weekend there is a
significant drop in the number of births. The bottom frame illustrates the
power spectrum of the detrended signal. It is surprising that the power
spectrum still shows a sign of a weekly component. The reason for this
unexpected effect is that the result is based on the implicit assumption
that the weekly component is the same throughout all the years: an incorrect
assumption. As a matter of fact, the practice of scheduling deliveries during the week has been increasing in prevalence since the 
late $1980$'s \cite{vitalstatistics}. Therefore we decided to evaluate the weekly component year by
year. We expect that in so doing no sign of weekly periodicity on the power
spectrum is left. Fig. \ref{weekperiodicity2} shows that this conjecture is
correct. In fact the bottom frame shows no signs of weekly periodicity. Note
that the ordinate scale is smaller than the scale of the previous figures by
a factor of 10. In the top frame we show the intensity of the weekly
component through the years. This intensity increases and this phenomenon
becomes especially significant in the last ten years.

\subsection{The Fluctuations}\label{ultimategoal} 
\begin{figure}[tbp]
\includegraphics[angle=-90,width=5.5in] {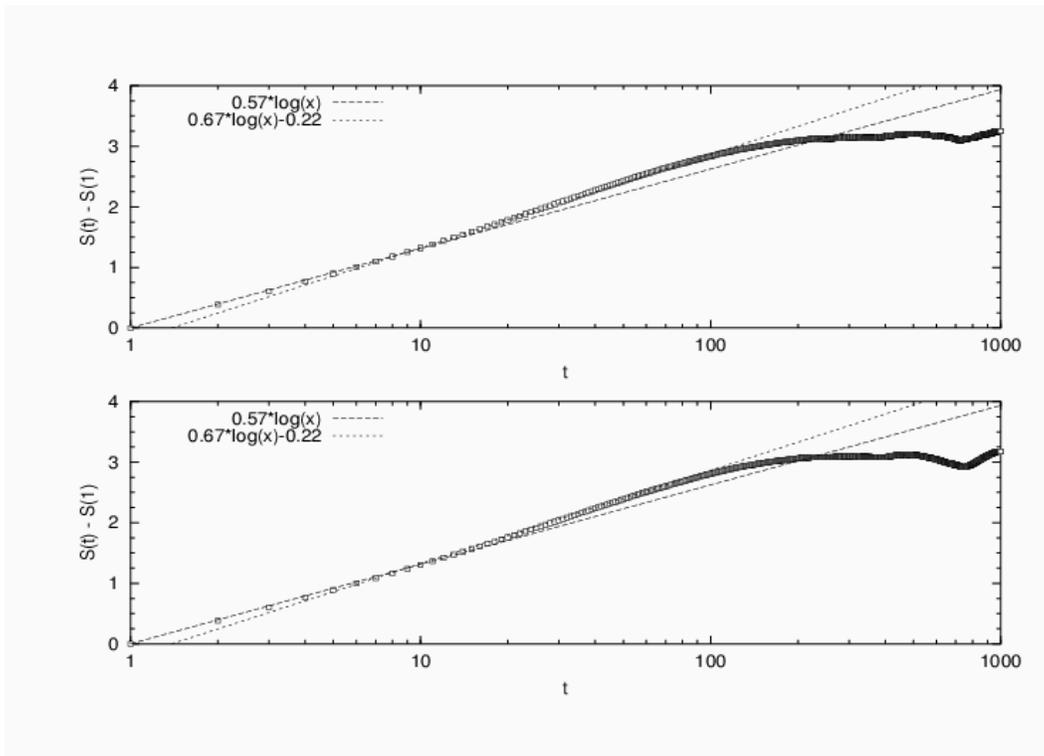}
\caption{The DE of the detrended data. The squares denotes the results of
the DE analysis on the data detrended with the step smoothing (top frame),
and wavelet smoothing (bottom frame).}
\label{denoise}
\end{figure}

\begin{figure}[tbp]
\includegraphics[angle=-90,width=5.5in] {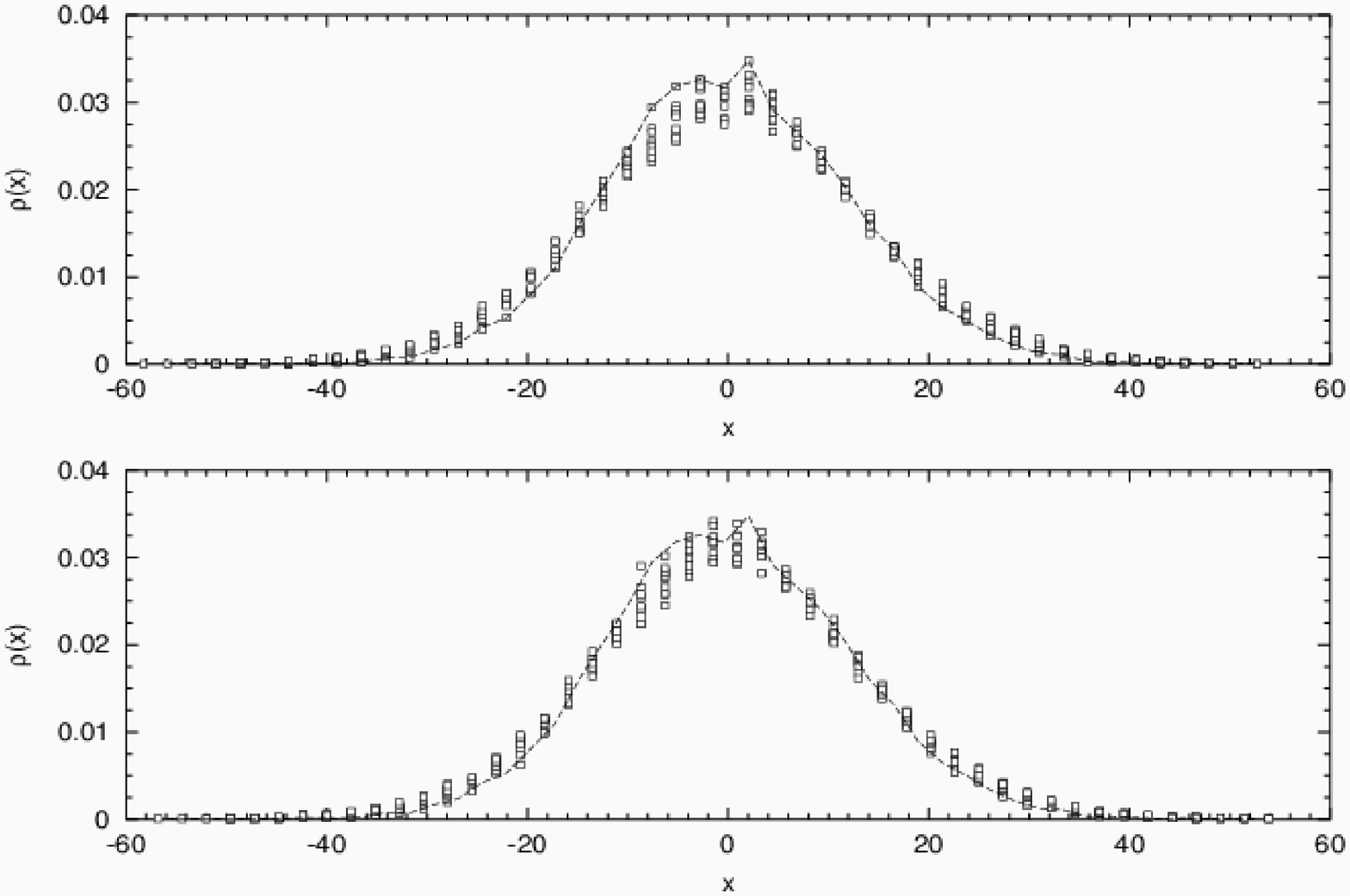}
\caption{The DAS at work in the time region from $1$ to $10$. The squares
denote the pdf rescaled with $\delta = 0.5$ for the data detrended with the
step smmothing ( top frame ) and with the wavelet smoothing ( bottom frame
), the dashed line correspond to the pdf at time $t=1$. The overlap between
squares and dashed line is not so good.}
\label{sqend05}
\end{figure}

\begin{figure}[tbp]
\includegraphics[angle=-90,width=5.5in] {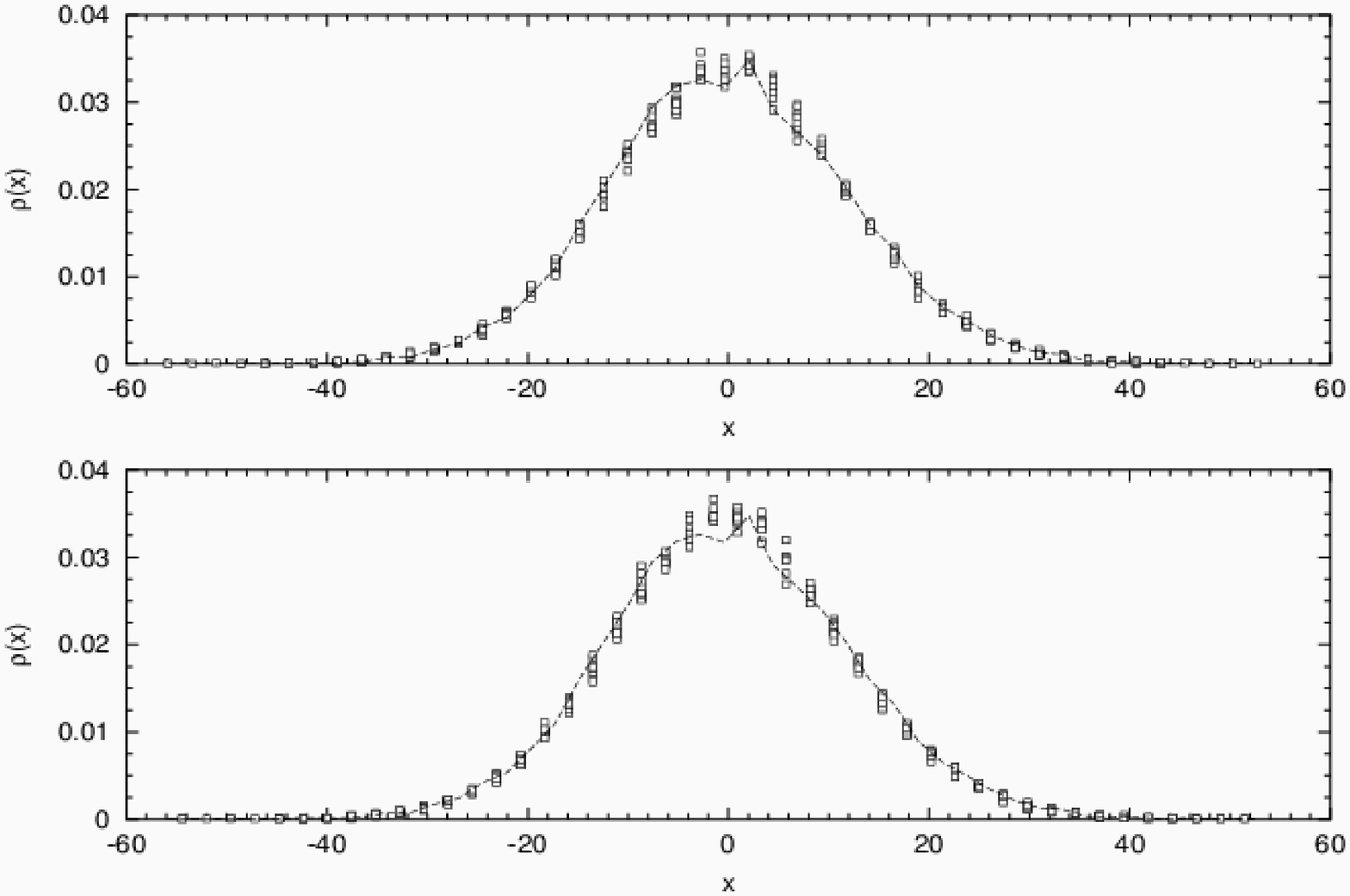}
\caption{The DAS at work in the time region from $1$ to $10$. The squares
denote the pdf rescaled with $\delta = 0.57$ for the data detrended with the
step smmothing (top frame) and with the wavelet smoothing (bottom frame), the dashed line correspond to the pdf at time $t=1$. The overlap between
squares and dashed line is good.}
\label{sqend057}
\end{figure}

\begin{figure}[tbp]
\includegraphics[angle=-90,width=5.5in] {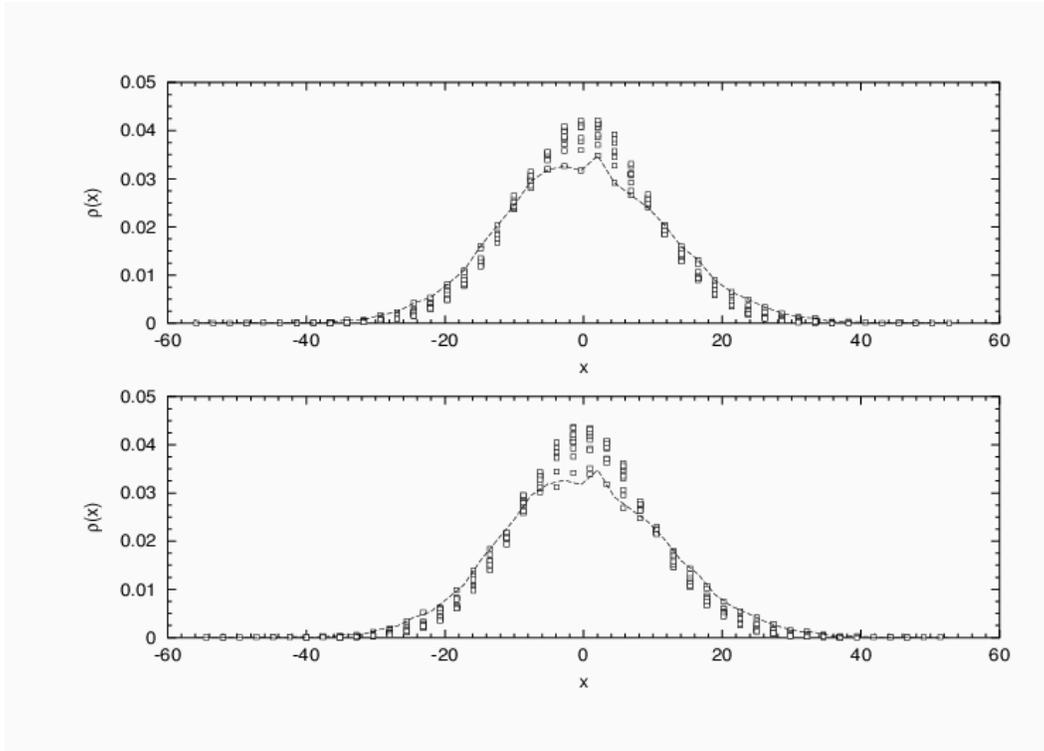}
\caption{The DAS at work in the time region from $1$ to $10$. The squares
denote the pdf rescaled with $\delta = 0.67$ for the data detrended with the
step smmothing (top frame) and with the wavelet smoothing (bottom frame), the dashed line correspond to the pdf at time $t=1$. The overlap between
squares and dashed line is bad in particular in the central part of the pdf.}
\label{sqend067}
\end{figure}

\begin{figure}[tbp]
\includegraphics[angle=-90,width=5.5in] {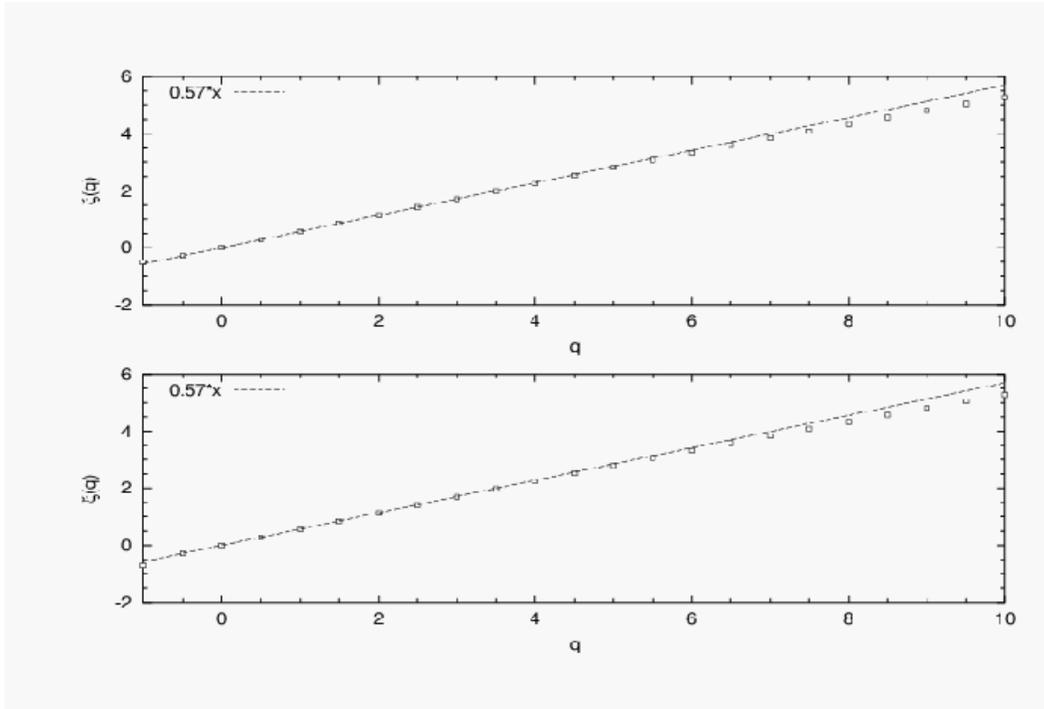}
\caption{The MS at work in the time region from $1$ to $10$. The squares
denote the results for the data detrended with the step smmothing (top
frame) and with the wavelet smoothing (bottom frame), the dashed line
correspond to a straight line of slope of $0.57$.}
\label{xdqres}
\end{figure}

\begin{figure}[tbp]
\includegraphics[angle=-90,width=5.5in] {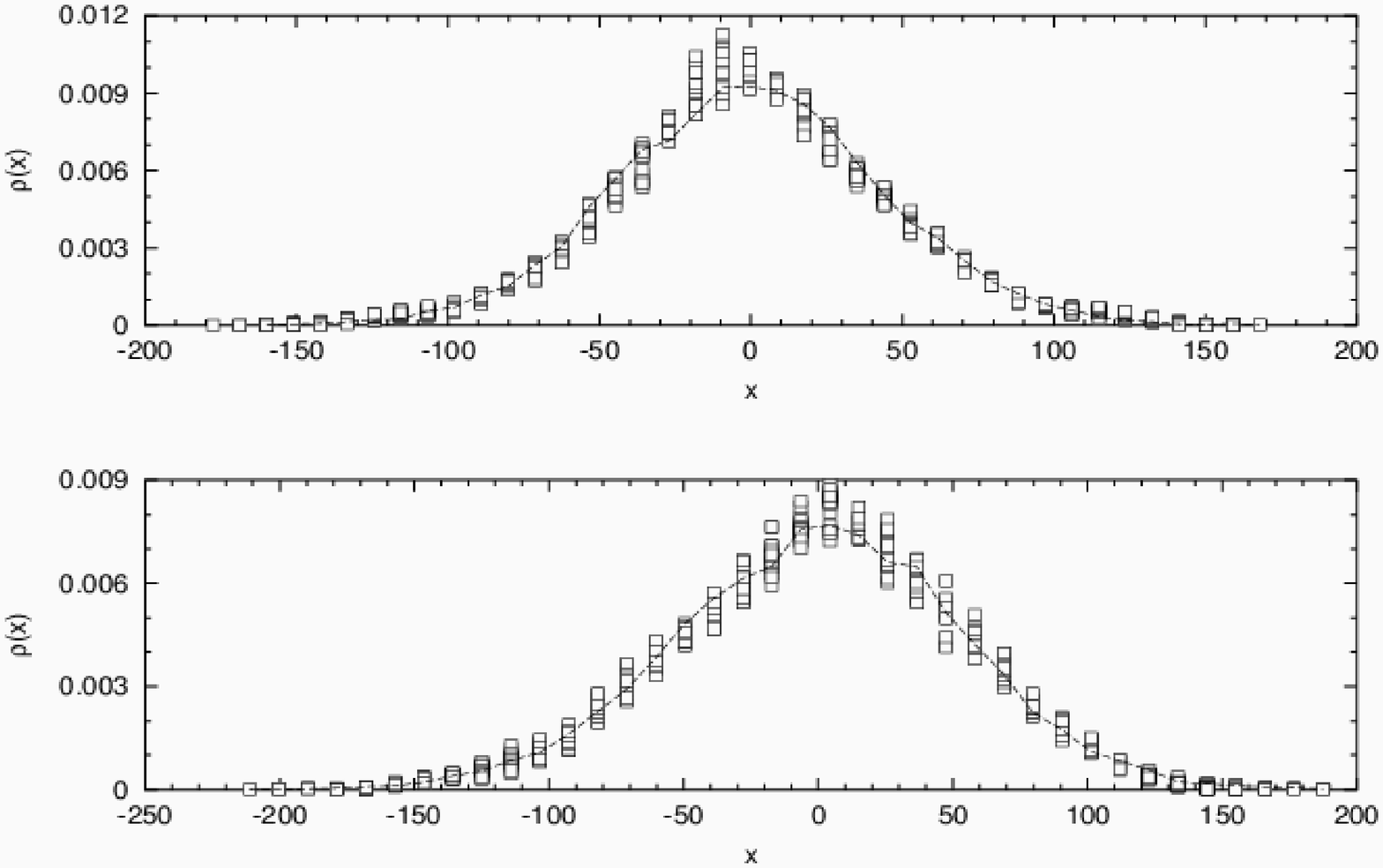}
\caption{The DAS at work in the time region from $10$ to $80$. The squares
denote the pdf rescaled with $\delta = 0.67$ for the data detrended with the
wavelet smmothing (top frame) and for an artificial data corresponding to
a FBM with $H=0.67$ (bottom frame) the dashed line correspond to the pdf
at time $t=10$.}
\label{testofsc}
\end{figure}

Finally, we apply the diffusion entropy analysis (DEA)
to the data with all possible forms of bias, seasonal and demographical,
detrended. Fig. \ref{denoise} shows that the first ten days are characterized
by $\delta _{de}=0.58$ and that immediately before the saturation regime,
caused by the detrending procedure, in the time region between $10$ and $80$%
, an even larger value of scaling index, $\delta _{de}=0.67$, emerges. Do
these parameters correspond to a proper realization of the scaling
definition of Eq. (1)? To answer this question we apply the DAS and the MS
to both time regions. In Figs. \ref{sqend05}, \ref{sqend057} and \ref
{sqend067} we show the DAS in the time region of the first ten days, with $%
\delta _{de}=0.5$, $\delta _{de}=0.57$ and $\delta _{de}=0.67$,
respectively. The same time region is analyzed in Fig. \ref{xdqres} by means
of the MS method. These figures indicate clearly that $\delta _{de}=0.57$
can be considered a genuine scaling parameter. The top frame of Fig. \ref
{testofsc} shows the results of the DAS in the region where $\delta
_{de}=0.67$. The result of the DAS applied to the time series stemming from
the step detrending procedure is virtually equivalent to that of the wavelet
procedure. For simplicity in Fig. \ref{testofsc} we report only the case of
the wavelet detrending procedure. We see that this $\delta _{de}$ does not
correspond to a satisfactory realization of Eq. (1).

On the basis of these results, we would be tempted to conclude that the
scaling $\delta =0.57$ is genuine and $\delta =0.67$ is not. However, to be
as rigorous as possible, we want to discuss first the intriguing issue of
the difference between \emph{real} and \emph{genuine} scaling. By real we
mean a scaling coresponding to a realization of Eq. (1). By genuine we mean
a scaling reflecting the cooperative properties of the process under study.
We cannot rule out the possibility that the scaling $\delta =0.67$ is real,
but not genuine. Let us see why. We notice that the length of our time
series is $13149$. We generate several artificial sequences, with the same
length, with the algorithimic prescription that Ref. \cite{feder} proposes
to build up fractional Brownian motion, with $H=0.67$. This corresponds to a
real scaling with $\delta =0.67$. Then we apply the DAS to the same time
region where the real data yield $\delta _{de}=0.67$. The results are
reported in the bottom frame of Fig. \ref{testofsc} and should be compared
to top frame, illustrating the analysis of the real data. We see that the
results of the artificial sequences as  ''good'' or as ''bad'' as those of
the real data. On the basis of this, we cannot dismiss the possibility that $%
\delta =0.67$ is essentially real scaling. 

Would the scaling also be genuine? Here we have to face two different
eventualities. The first is that the genuine fluctuations remaining after
detrending are a generalization of the dynamic model proposed years ago in
Ref. \cite{cmm}. This model, the Copying Mistake Map (CMM), assumes that the
time series is generated by a composite of two mechanisms. The first,
adopted with larger probability, is a prescription generating uncorrelated
fluctuations, and the second, applying with a much smaller probability, is a
prescription generating correlated fluctuations, and consequently a
diffusion process faster than the uncorrelated component. In the large time
scale regime the second component dominates the diffusion process, thereby
producing a crossover from normal to anomalous scaling. If we replace the
random component with fluctuations characterized by anomalous scaling,
weaker than the second component, we expect a crossover from a scaling
larger than the ordinary to an even larger scaling. This might be the model
behind the results illustrated in Fig. \ref{testofsc}. In this case the scaling $\delta
=0.67$ would be genuine as well as real. However, we canot rule out the
possibility that the effect is not genuine. This effect might be due to the
presence of a residual contribution of the slow component that would
generate in the long-time regime a difffusion faster than that produced by
the correlation stemming from the genuine complexity of the process under
study, in the same way as the generalized CMM model would do. However, in
this case the effect might be real or not, but it would not be genuine.

With all these caveates in mind, we can reach the following conclusion. The
scaling $\delta =0.57$ is both real and genuine. At the present time we
cannot reach a conclusion about the scaling $\delta =0.67$.

\section{Concluding Remarks}\label{conclusion}
The most important result of this paper is the detection, through the introduction of the 
model of Eq. (\ref{teenmodel0}) and the use of the detrending techniques of \cite{paper1}, 
of the genuine scaling $\delta =0.57$. There are several indications that
this is the real complexity strength of the process. First, the adoption of
two distinct detrending techniques yields the same slow-motion component.
Second, the study of artificial time series with this same kind of
slow-motion component explains with sufficient clarity that the detrending
procedure yields a saturation effect, and proves that the time region before
this saturation effect can be safely used to detect the genuine scaling. Third the value 
$\delta = 0.57$ has been tested with multifractal test and the DAS analysis.

Finally we want to stress again the important difference between real and real and genuine scaling. This  is so  because the application of nonlinear dynamical analysis to sociological problems holds promise for better understanding the complex processes that are responsible for phenomena such as teen pregnancy and birth.  However, extracting accurate and useful information from observable events such as births to teens, measured and analyzed in the form of a time series, is quite difficult. This difficulty arises from the fact that complex cultural, biological and sociological interactions are embedded within more periodic cycles of events.  Thus, their appearance in the time series is masked in such a way as to make their accurate identification and quantification major challenges for investigators. One strategy used to disentangle and isolate complexity for quantification is simply to remove all major, identifiable periodicities from the data, and to assume that the portion of what remains that cannot be described as ordinary diffusion can be labeled �enuine�complexity.  
This paper has shown that such a strategy is not without shortcomings which increase the danger of over-estimating the genuine complexity of the sociological process.  Over-estimating the genuine complexity of any process leads to errors in constructing a useful theoretical model for describing, understanding, predicting and controlling the process.           

Acknowledgment: MI and PG thankfully acknowledge ARO for financial support
through grant DAAD19-02-0037. PH acknowledge support from NICHD grant R03.

\newpage

\begin{references}


\bibitem{paper1} M. Ignaccolo, P. Allegrini, P. Grigolini, P. Hamilton and B. J. West, submitted to Physica A.

\bibitem{stanley} C.-K. Peng, S.V. Buldyrev, S. Havlin, M. Simons, H.E. Stanley, and A.L. Goldberger, Phys. Rev. E {\bf 49}, 1685 (1994).

\bibitem{patti} N. Scafetta, P. Hamilton, P. Grigolini and B.J.West, Phys.  (condmat) 0208117

\bibitem{zeromean} We consider the components $\Phi_{j}^{week}$ and $\Phi_{j}^{year}$ to have a null sum over one period, namely 
$\sum\limits_{j=1}^{7} \Phi_{j}^{week} = 0$ and $\sum\limits_{j=1}^{365} \Phi_{j}^{year} = 0$.

\bibitem{forgetaverage} We disregard the contribution to the sum of Eq. (\ref{annmovingaverage}), due to the average of the teen birth data $S$, since this consists, trivially, in a translation.

\bibitem{vitalstatistics} S. Curtin, M.A. Park and  M. Park, \textit{Trends in the attendance, place and timing of births, and in the use of obstetric interventions: United States, 1989-1999}, National Vital Statistics Reports , Vol. {\bf47}, Num. 27 (1999)

\bibitem{feder} J. Feder, \emph{fractals} Plenum Press, New York (1988).

\bibitem{cmm} P. Allegrini, M. Barbi,P. Grigolini and B.J. West, Phys. Rev. E {\bf52}, 5281 (1995). 

\end{references}
\end{document}